\documentclass[12pt]{article}
\usepackage{amsmath}
\usepackage{amsfonts}
\usepackage{appendix}
\begin{document}
\title{Massless Fermions in anisotropic Bianchi type I spacetimes}
\author{Matthias Wollensak\footnote{matthias.wollensak@uni-jena.de} \\Theoretisch - Physikalisches Institut,
\\Friedrich - Schiller - Universit$\ddot{a}$t Jena,
\\Max - Wien - Platz 1, D - 07743 Jena, Germany}
\date{21. April 2016}
\maketitle

\begin{abstract}
The behavior of spin - 1/2 - particles in anisotropic Bianchi type I backgrounds is investigated utilizing the concept of differential forms and orthonormal frames. Specializing to the massless case and power law scale factors $\alpha_j(t) = t^{q_j}$ of the metric where $q_1 = q_2 $, an analytical outcome for the time evolution operator in terms of Bessel functions is presented.
\end{abstract}

\section*{I. INTRODUCTION}
The behavior of particles in a curved background which obey the Dirac equation is of considerable interest in cosmology and astrophysics. An early discussion of the massless case has been given by Brill and Wheeler [1]. An investigation of quantized massive and massless spin - 1/2 fields in conformally flat Robertson - Walker (RW) universes has been performed by Parker who showed that no production of massless particles is taking place in those spacetime [2].

While it is comparatively easy to find exact classical solutions of Dirac`s equation in some isotropic backgrounds this is unfortunately not the case for anisotropic backgrounds, even if one considers only massless fields. Apart from a few exactly solvable models such as for example the stiff - matter model treated in sect. III. one must therefore usually resort to appropriate approximations. In an approach utilized by Zel`dovich and Starobinsky one considers e. g. small anisotropic perturbations about a RW spacetime [3].

In this paper we derive the time evolution operator for Weyl - spinors which is nonperturbative with respect to the underlying spacetime, where we focus on anisotropic backgrounds described by Bianchi type I (BI) spacetimes. These spacetimes are a natural generalization of flat RW spacetimes to anisotropic spacetimes where the time dependence of the scale factors for each of the spatial directions is no longer the same. Although the present day universe seems to a very high degree to be isotropic, this needed not necessarily be the case at a very early stage of development of the universe. In fact it has been suggested by Zel`dovich that starting with an anisotropic universe and taking into account quantum effects in the vicinity of the initial singularity could lead to an isotropization of the universe at the Planck time due to particle creation processes [4]. Recently, the issue of anisotropic BI spacetimes has been discussed in the context of preinflationary scenarios of the universe [5].

Classical solutions of the Dirac equation in BI spacetimes have been investigated e.g. by Henneaux who considered gravitational and spinor fields being both invariant under an Abelian 3 - parameter group of transformations [6], or by Saha and Boyadjiev who studied (interacting though space - independent) spinor and scalar fields [7].

The structure of this paper is as follows: In sect. II. we set up the Dirac equation in BI backgrounds with respect to orthonormal frames and specialize to the case of massless fermions in planar BI spacetimes with power law scale factors. We derive for those backgrounds an approximate expression of the time evolution operator. This solution exhibits for a wide range of parameters of these spacetimes the correct asymptotic behavior and the exact short - time limit as well. Moreover,  it comprises the exact flat RW solution as a limiting case, which is explicitly demonstrated in sect. III. We also treat in this sect. an anisotropic so-called stiff matter model being one of those rare exactly soluble cases, and show that our solution agrees with the exact result for small times and in the asymptotic limit.
\\
\\
\\
\\
\\
\section*{II. MASSLESS FERMIONS IN PLANAR BI SPACETIMES}
We start with the formulation of Dirac`s equation using the differential geometric approach.\footnote{An analogous treatment of Maxwell`s equations has been performed in ref. [8]} Usually the line element of a Bianchi type I universe is defined by

\begin{equation}
ds^2 = dt^2 - \sum\limits_{i=1}^{3} \alpha_{i}^{2} (t)(dx^{i})^2
\end{equation}

with metric tensor $g = g_{\mu\nu} dx^{\mu} \otimes dx^{\nu}$ where $g_{\mu\nu} = diag(1,- \alpha_{1}^{2},- \alpha_{2}^{2},- \alpha_{3}^{2})$. An orthonormal frame $\{\Theta^{\mu} \}_{\mu}$ with elements $\Theta^{\mu}$ in the cotangent space is e.g. given by the following covectors $\Theta^0 = dt,\ \Theta^j = \alpha_j(t) dx^j$, whereas $e_0 = \partial_t, \  e_j = \alpha_j^ {-1}(t) \ \partial_j$ (j = 1, 2, 3) represent the corresponding basis vectorfields of the tangent space. The line element (1) reads in this case $ds^2 = (\Theta^0)^2 - (\Theta^1)^2 - (\Theta^2)^2 - (\Theta^3)^2$, and the metric tensor takes the particularly simple form $g = \eta_{\mu\nu} \Theta^{\mu} \otimes \Theta^{\nu}$ with $\eta_{\mu\nu} = g(e_{\mu} , e_{\nu})  \equiv  diag(1, -1,-1,-1) $, since $\Theta^{\mu}(e_{\nu}) = \delta^{\mu}_{\nu}$ [9], [10]. The exterior derivative of the covector $\Theta^j$ is given by the 2-form

\begin{equation}
d\Theta^j = - C^j_ {\ 0j} \Theta^0 \wedge \Theta^j
\end{equation}

(no summation!) where the nonvanishing commutation coefficients $C^{\alpha}_ {\ \mu \nu}$ read

\begin{equation}
C^j_ {\ 0j} = - \frac{\dot{\alpha_j}}{\alpha_j}.
\end{equation}

Since in general orthonormal frames are not necessarily holonomic, the exterior derivative of covectors and correspondingly the commutator of vectors do not vanish any more. Instead holds now $d\Theta^{\mu} = - \frac{1}{2} C^{\mu}_ {\ \alpha \beta} \Theta^{\alpha} \wedge \Theta^{\beta}$ and $[e_{\mu}, e_{\nu}] = C^{\alpha}_ {\ \mu \nu} e_{\alpha}$, with $C^{\alpha}_ {\ \mu \nu} = -C^{\alpha}_ {\ \nu \mu}$. There are three nonvanishing entries of the connection 1 - form, $\omega^j_ {\ 0}  =  C^j_ {\ 0j}  \Theta^j$ (no summation!), and since the covariant differentiation of a Dirac bispinor is given by the 1-form [10]

\begin{equation}
(D\psi)^{I} = d\psi^{I} + \frac{1}{8} \ \omega^{\mu}_ {\ \nu} \ [\gamma_{\mu}, \gamma^{\nu}]^{I}_{\ J} \psi^{J} ,
\end{equation}

the Dirac equation takes the form

\begin{equation}
\left\lbrace e_0 - \gamma_0 \sum\limits_{j=1}^{3}
\gamma_j e_j  - \frac{1}{2} \, {\cal{C}} + i \gamma_0 m \right\rbrace \psi = 0 .
\end{equation}

Here, ${\cal{C}} := \sum^3_{j=1} C^j_ {\ 0j} = - \partial_t \, ln(\sqrt{|g|})$ with $g = det g_{\mu \nu}$, and the $\gamma^{\mu}$ are the flat spacetime Dirac matrices. The spatial translation invariance of spacetime (1) motivates the ansatz

\begin{equation}
\psi(\textbf{x},t) = c_\textbf{k} \, e^{i\textbf{k} \textbf{x}} \left( \begin{array}{c} \varphi(\textbf{k}, t)  \\ \chi(\textbf{k}, t)  \end{array} \right) 
\end{equation}

with normalization constant $c_\textbf{k}$, and $\varphi, \chi$ denote two-component spinors transforming according to the representations $(\frac{1}{2}, 0)$ and $(0, \frac{1}{2})$. Eq. (5) can be rearranged as a coupled system for those spinors:

\begin{equation}
\begin{aligned}
e_0 \left(e^{+ imt} |g|^{1/4 } \varphi(\textbf{k}, t) \right) - i \textbf{p} \boldsymbol\sigma \, e^{+ imt} |g|^{ 1/4 } \chi(\textbf{k}, t) & = 0  
\\
e_0 \left(e^{- imt} |g|^{ 1/4 } \chi(\textbf{k}, t) \right) - i \textbf{p} \boldsymbol\sigma \, e^{- imt} |g|^{ 1/4} \varphi(\textbf{k}, t)  &= 0
\end{aligned}
\end{equation}

where the physical 3 - momentum $\textbf{p}$ is

\begin{equation}
p_j(t) = \frac{k_j}{\alpha_j(t)} .
\end{equation}

In the massless case the situation simplifies because the decoupling of the above system of differential equations can be easily achieved by setting $\chi = -\, \varphi$ and $\chi = +\, \varphi$ as well. The first case defines the eigenspinor of the chirality operator $\gamma_5$ with eigenvalue -1, while the latter belongs to the eigenspinor with eigenvalue +1. Hence, in ansatz (6) the bispinor $\psi$ can be labeled by $\mp$, i.e. $\psi^{\mp} \sim (\varphi^{(\mp)}, \mp \varphi^{(\mp)})^T$, referring to an eigenstate of $\gamma_5$ with negative and positive chirality, resp.

We restrict ourselves from now on to the massless case and introduce the Weyl - spinors $\phi^{(\mp)}$ with components

\begin{equation}
\phi_J^{(\mp)}(\textbf{k}, t) = |g(t)|^{1/4} \, \exp \left( \pm (- 1)^J \, i \int \limits_{t_{\tilde{A}}}^{t} p_3(x) dx  \right)  \, \varphi_J^{(\mp)}(\textbf{k}, t)|  .
\end{equation}

Then one obtains from (7) the following system:

\begin{equation}
\partial_t \phi^{(j, \mp)}(\textbf{k}, t) - \Omega^{(\mp)}(\textbf{k}, t) \phi^{(j, \mp)}(\textbf{k}, t) = 0 \ \ (j = 1, 2)
\end{equation}

where the two linearly independent solutions for each chirality state are labeled by j, and where the matrix $\Omega^{(\mp)}$ is defined to be

\begin{equation}
\Omega^{(\mp)}(\textbf{k}, t)  = \left( \begin{array}{rr} 0 \  \  \  \  \  \  \  \    &  \gamma^{(\mp)}(\textbf{k}, t)  \\   - [\gamma^{(\mp)}(\textbf{k}, t)]^{\ast}  &  0 \  \  \  \  \  \   \end{array} \right)
\end{equation}

with

\begin{equation}
\gamma^{(\mp)}(\textbf{k}, t)  = \pm (ip_1 + p_2)  \exp \left( \mp 2i \int \limits_{t_{\tilde{A}}}^{t} p_3(x) dx  \right)
\end{equation}

($t_{\tilde{A}} \geq 0$). On solving (10) one can determine the pertaining four bispinor solutions satisfying (5) by virtue of (6) and (9):

\begin{equation}
\psi^{(j, \mp)}(\textbf{x},t) = c^{(j, \mp)}_\textbf{k} \, e^{i\textbf{k} \textbf{x}} \left( \begin{array}{c} \varphi^{(j, \mp)}(\textbf{k}, t)  \\ \mp \, \varphi^{(j, \mp)}(\textbf{k}, t)  \end{array} \right) .
\end{equation}

$(j  = 1, 2)$. Eq. (12) implies $\gamma^{(-)}(\textbf{k}, t) \rightarrow \gamma^{(+)}(\textbf{k}, t)$ when $\textbf{k} \rightarrow  - \textbf{k}$. Hence one must only find the two solutions of, say, the negative chirality case and gets the positive chirality solutions according to

\begin{equation}
\phi^{(j, +)}(\textbf{k}, t) = \phi^{(j, -)}(- \textbf{k}, t).
\end{equation}
\\
An appropriate scalar product is defined by

\begin{equation}
\left\langle u,v \right\rangle = \int\limits_{\Sigma} \ast F_{u,v}
\end{equation}

with $\Sigma$ a spacelike Cauchy hypersurface and $\ast$ the duality operator. Furthermore, $F_{u,v}$ denotes a 1-form given by

\begin{equation}
F_{u,v}(x) = (F_{u,v}(x))_{\mu} \, \Theta^{\mu} := \overline{u(x)} \,  \gamma_{\mu} v(x) \,\Theta^{\mu} ,
\end{equation}

where $u$ and $v$ are solutions of (5), and $\overline{u}$ stands for the Dirac adjoint. We specialize now to spacetimes (1) with $\alpha_1(t) = \alpha_2(t) = t^{\nu}, \ \alpha_3(t) = t^{1 - \mu} \ ( \mu >  0)$, i.e.

\begin{equation}
ds^2 = dt^2 -  t^{2 \nu} (dx^1)^2 -  t^{2 \nu} (dx^2)^2 -  t^{2 - 2 \mu} (dx^3)^2.
\end{equation}

The components of the 3 - momentum $\textbf{p}$ are $p_1 = k_1t^{- \nu}, p_2 = k_2t^{- \nu}$ and $p_3 = k_3t^{\mu - 1}$. On defining

\begin{equation}
k_{\pm} = (k_2 \pm ik_1) \, \exp \left(\pm \frac{ 2ik_3}{\mu} \, t^{\mu}_{\tilde{A}}\right)
\end{equation}

one gets from (12)

\begin{equation}
\gamma^{(-)}(\textbf{k}, t) = \frac{k_+}{t^{\nu}} \exp \left( \frac{- 2ik_3}{\mu} \, t^{\mu} \right).
\end{equation}

In what follows we seek negative chirality solutions of equation (10). These solutions can be formally written as

\begin{equation}
\begin{aligned}
\phi^{(-)}(\textbf{k}, t) \, = \, \lim_{n \to \infty} \left( \hat{\Omega}_{\textbf{k}}^n[\theta] \right) (t) \, = \,   [ \, 1 + \sum_{n=1}^{\infty} &  \int\limits_{t_A}^t dt_1 \int\limits_{t_A}^{t_1} dt_2 ...\int\limits_{t_A}^{t_{n-1}} dt_n \, \Omega^{(-)}(\textbf{k}, t_1)
\\
&\times \Omega^{(-)}(\textbf{k}, t_2) ... \Omega^{(-)}(\textbf{k}, t_n) \, ] \ \theta_A(\textbf{k}).
\end{aligned}
\end{equation}

It is useful to introduce the new time variable

\begin{equation}
s(t) = t^{\mu}
\end{equation}

and the parameter

\begin{equation}
\delta = \frac{1 - \nu}{\mu}.
\end{equation}

Thus one gets (we conveniently choose $\theta_A(\textbf{k}) \equiv \phi^{(-)}(\textbf{k}, t_A)$) 

\begin{equation}
\phi^{(-)}(\textbf{k} ,t)  = K_{\textbf{k}}^{(-)}(t|t_A) \, \phi^{(-)}(\textbf{k}, t_A)
\end{equation}

with the time evolution operator for the negative chirality spinors given by

\begin{equation}
K_{\textbf{k}}^{(-)}(t|t_A)  = 1 + \sum_{n=1}^{\infty} \left( \begin{array}{rr}  I_n(s;1)  &   0 \ \ \    \\   0 \ \ \  &  I^{\ast}_n(s;1)   \end{array} \right) 
\left( \begin{array}{rr} 0 \ \ \   &  \frac{k_2 + ik_1}{\kappa}  \\   - \frac{k_2 - ik_1}{\kappa}  &  0 \ \ \   \end{array} \right)^n
\end{equation}

where

\begin{equation}
\begin{aligned}
I_n(s;y)  = \left(  \frac{2 \eta_{ \delta}}{\mu^2} \right)^{n/2}  (|k_3| s)^{\delta n} \int\limits_{\sigma_A}^y d\sigma_1 \int\limits_{\sigma_A}^{\sigma_1} d\sigma_2  ...\int\limits_{\sigma_A}^{\sigma_{n-1}}  &  d\sigma_n 
\left[ \prod\limits_{l = 1}^{n} \sigma_l^{\delta -1} \right]
\\
& \times
\exp\left[ i\frac{2k_3s}{\mu} \sum\limits_{m = 1}^{n} (- 1)^m \sigma_m  \right].
\end{aligned}
\end{equation}

Here, we defined

\begin{equation}
\kappa := \sqrt{k^2_1 + k^2_2}
\end{equation}

\begin{equation}
\eta_{\delta} := \frac{\kappa^2}{2 |k_3|^{2 \delta}}
\end{equation}

and

\begin{equation}
\sigma_j(t) := \frac{s_j(t)}{s(t)}, \ \ \sigma_A(t) := \frac{s(t_A)}{s(t)}.
\end{equation}

Note that $t_A = 0$ enforces $\delta > 0$ (i.e. $\nu < 1$), whereas $\delta$ can also be negative (i.e. $\nu \geq 1$) provided $t_A > 0$. Equations (23), (24) and (25) together with (13), (9) represent the two linearly independent exact negative chirality solutions of a massless spin - $\frac{1}{2}$ - field in spacetimes (17). However, they are not very illuminating as far as the explicit behavior of fermions propagating in backgrounds described by (17) is concerned. On the other hand, the form of $I_n(s;y)$ suggests the following approximation: we first put $\epsilon_l = 1 - \sigma_l$ implying $0 \leq \epsilon_1 \leq ...\leq \epsilon_{n - 1} \leq \epsilon_n \leq 1$ and write $\sigma_l^{\delta - 1}  \approx e^{(1 - \delta) \epsilon_l}$ for $l < n$, i. e. we keep the  factor $ \sigma_n^{\delta - 1}$ (the range of the parameter $\delta$ is restricted from now on by $\delta \leq 1$), so that

\begin{equation}
\prod\limits_{l = 1}^{n} \sigma_l^{\delta -1} \approx  e^{(1 - \delta) (n - 1)} \exp \left[(\delta - 1) \sum\limits_{l = 1}^{n - 1} \sigma_l \right]   \sigma_n^{\delta - 1} .
\end{equation}

On substituting (29) into (25) one then obtains 

\begin{equation}
\begin{aligned}
I_n(s;y) \approx \left(  \frac{2 \eta_{ \delta}}{\mu^2} \right)^{n/2}  (|k_3| s)^{\delta n} &\  e^{(1 - \delta) (n - 1)} \int\limits_{\sigma_A}^y d\sigma_1 \int\limits_{\sigma_A}^{\sigma_1} d\sigma_2 ...\int\limits_{\sigma_A}^{\sigma_{n-1}} d\sigma_n \sigma_n^{\delta -1}
\\
& \times \exp \left[  (\delta -1) \sum\limits_{l = 1}^{n - 1} \sigma_l + i \frac{2k_3s}{\mu} \sum\limits_{m = 1}^{n} (- 1)^m \sigma_m \right] .
\end{aligned}
\end{equation}

This result yields the correct short - time 
limit.\footnote{Note that $\sigma_A \rightarrow  1$ when $t  \rightarrow  t_A$, so that $\epsilon_l  \ll 1$ is obviously satisfied for $I_n(s;1)$. But one still gets the correct behavior for small times even when one puts from the onset $\sigma_A = 0$!}
Approximation (29) is, however, still sensible in the asymptotic case despite the fact that $\epsilon_l  \ll 1$ (for $l < n$)  is no longer guaranteed when $|k_3| s \rightarrow \infty$, because the lower integration limits $\sigma_A$ tend to 
zero.\footnote{It should be emphasized here that (29) does not imply any restriction on $ \epsilon_n \,\in [0; 1] !$}
 For consider the "worst case" $\sigma_l \ll 1$ and $\epsilon_l \approx 1$, resp. ($l = 1, 2, ... ,n$), which is equivalent to $\sigma_1 \ll 1$ ($\epsilon_1 \approx 1$). It is useful to define the positive quantity $\Sigma_{l;n} := \sum_{m=l}^{n} (- 1)^{m + l} \sigma_m$ satisfying $\sigma_l \geq   \Sigma_{l;n} \geq  \sigma_l - \sigma_{l+1}$ and to introduce a new very small positive parameter $\epsilon = O(|k_3 s|^{-1})$ ($\epsilon > \sigma_A$). Then the phase factor $\exp [- i \, 2k_3 s \, \Sigma_{1;n} / \mu]$ in the integrand of $I_n(s;\epsilon)$ is basically independent of any changes $\Delta \sigma_m$ and can therefore be put in front of the multiple integral. As a consequence, $(2 \eta_{ \delta}/ \mu^2 )^{-n/2}  (|k_3| s)^{- \delta n} I_n(s;\epsilon)$ can be estimated,  apart from this phase factor, by $\epsilon^{\delta n} / n! \delta^n$ in the exact case (25) and by $e^{(1 - \delta) (n - 1)}  \epsilon^{\delta + n - 1} [1 + O((n - 1)\, \epsilon )]\,  \Gamma(\delta) / \Gamma(\delta + n)$ in the approximate case (30). It follows that (30) exhibits qualitatively the same behavior as (25) and even matches it in the most dominant case $n = 1 $, so that the diagonal entries on the r. h. s. of (24) read in this case $1 +  O(\epsilon^{\delta + 1}) [1 + O(\epsilon^2)]$ while the off - diagonal entries are given by $O(\epsilon^{\delta}) [1 + O(\epsilon^2)]$, whereas with (25) one gets for the diagonal elements $1 +  O(\epsilon^{2 \delta}) [1 + O(\epsilon^{2 \delta})]$ and the off - diagonal elements are described by $O(\epsilon^{\delta}) [1 + O(\epsilon^{2 \delta})]$. 

In order to proceed we rewrite the r.h.s. of  eq. (30) as a Laplace convolution:

\begin{equation}
\begin{aligned}
I_n(s) \approx  \left(  \frac{2 \eta_{ \delta}}{\mu^2} \right)^{n/2}&  (|k_3| s)^{\delta n} \ e^{- i\frac{k_3 s}{\mu}}
\\
& \times \int\limits_{0}^{1 - \sigma_A}  dz \, h_n(1 - \sigma_A - z)
\left( g_{0, n - 1}  \ast g_{1, n - 1}  \ast ... \ast g_{n - 1, n - 1}  \right) (z)
\end{aligned}
\end{equation}

with

\begin{equation}
\begin{aligned}
g_{l, j}(z) & = \exp \left\lbrace \left[ (1 - \delta)(j - l) + i (- 1)^{l} \, \frac{k_3 s}{\mu}   \right] z \right\rbrace
\\
h_j(z)  & = (z + \sigma_A)^{\delta - 1} \exp \left\lbrace   i (- 1)^{j} \, \frac{k_3 s}{\mu}   (z + \sigma_A) \right\rbrace .
\end{aligned}
\end{equation}

Here, the asterisk denotes the usual Laplace convolution product. By use of Laplace transformation techniques the convolution in eq. (31) can be readily computed:

\begin{equation}
\left( g_{0, n - 1}  \ast g_{1, n - 1}  \ast ... \ast g_{n - 1, n - 1}  \right) (z) \ = \
 \sum\limits_{l = 1}^{n} \frac{e^{{\cal{G}}_{l, n}(s) \, z}}{{\prod\limits_{m = 1}^{n}}^{\prime}  \, \Delta {\cal{G}}_{l, m}(s)}  
\end{equation}

where $\Delta {\cal{G}}_{l, m}(s) := {\cal{G}}_{l, j}(s) - {\cal{G}}_{m, j}(s)$ and ${\cal{G}}_{l, j}(s) := - i (- 1)^{l} \, k_3 s/\mu + (1 - \delta)(j - l).$ The prime in the product on the r.h.s. of eq. (33) indicates that the factor $\Delta {\cal{G}}_{l, l}(s) = 0$ has to be omitted. Inserting eq.s (31) and (33) into (24) and performing a somewhat tedious calculation which consists basically in rearranging terms in such a manner that the ocurring summations can be explicitly carried out, one finally gets the following expressions for the matrix elements of the time evolution operator

\begin{equation}
\begin{aligned}
\left( K_{\textbf{k}}^{(-)}(t|t_A) \right)_{11} & = 1 + i (1 - \delta) \, x(s) \int\limits_0^{1 - \sigma_A(s)} dz \, (1 - z)^{\delta - 1} \, V_{\textbf{k}}(z; s)
\\
\left( K_{\textbf{k}}^{(-)}(t|t_A) \right)_{22} &= \left( K_{\textbf{k}}^{(-)}(t|t_A) \right)_{11}^{\ast}
\\
\left( K_{\textbf{k}}^{(-)}(t|t_A) \right)_{12} & =  \frac{k_+}{\kappa} (1 - \delta) \, x(s) \, e^{- 2i k_3 s/\mu} \int\limits_0^{1 - \sigma_A(s)} dz \, (1 - z)^{\delta - 1} \, U_{\textbf{k}}(z; s)
\\
\left( K_{\textbf{k}}^{(-)}(t|t_A) \right)_{21} &= -  \left( K_{\textbf{k}}^{(-)}(t|t_A) \right)_{12}^{\ast}
\end{aligned}
\end{equation}

with

\begin{equation}
V_{\textbf{k}}(z; s) = i \, e^{(1 - \delta) \lambda^{\ast}(s) z}\ \frac{R(z; s)}{Z(0; s)}
\end{equation}

and

\begin{equation}
U_{\textbf{k}}(z; s) =  e^{(1 - \delta) \lambda(s) z}\ \frac{Z(z; s)}{Z(0; s)} .
\end{equation}

where

\begin{equation}
R(z; s) = J_{- \lambda(s)}(x(s)) J_{\lambda(s)} [x(s) e^{(1 - \delta) z}] - J_{\lambda(s)}(x(s)) J_{- \lambda(s)}[x(s) e^{(1 - \delta) z}]
\end{equation}

\begin{equation}
Z(z; s) =  J_{- \lambda(s)}(x(s)) J_{- \lambda^{\ast}(s)} [x(s) e^{(1 - \delta) z}] + J_{\lambda(s)}(x(s)) J_{\lambda^{\ast}(s)}[x(s) e^{(1 - \delta) z}].
\end{equation}

Here, $x$ and $\lambda$ are given by

\begin{equation}
\begin{aligned}
x(s) &= \frac{ \sqrt{2 \eta_{ \delta}} \,  (|k_3| s)^{\delta}}{\mu (1 - \delta)}
\\
\lambda(s) &= \frac{1}{2} + i\, \frac{k_3 s}{\mu (1 - \delta)},
\end{aligned}
\end{equation}

and $J_{\lambda}(x)$ denotes Bessel`s function of the first kind [11].
\\
\\
\\
\\
\\
\section*{III. COMPARISON WITH EXACT RESULTS}
We first consider massless spin - $\frac{1}{2}$ - particles in a radiation - dominated universe with line - element (17) where $\mu = 1/2, \ \  \nu = 1/2$ (that is with (22): $\delta = 1$), a special case of a flat RW universe. In this case our result (34) represents the exact solution, which is actually true for all spacetimes (17) satisfying the condition $\delta = 1$.

On performing a $1 - \delta$ - expansion of  eq.s (34) one obtains for $\mu = \nu = 1/2$ the matrix elements (in the limit $\delta  \rightarrow  1$):

\begin{equation}
\begin{aligned}
\left( K_{\textbf{k}}^{(-)}(t|t_A) \right)_{11} &= e^{- 2 i k_3 (\sqrt{t} - \sqrt{t_A} ) } \left\lbrace 
\cos[2k (\sqrt{t} - \sqrt{t_A})] + i \, \frac{k_3}{k} \, \sin[2k (\sqrt{t} - \sqrt{t_A})]  \right\rbrace
\\
\left( K_{\textbf{k}}^{(-)}(t|t_A) \right)_{12} &= \frac{ k_+(t_{\tilde{A}})}{k} \, 
e^{- 2 i k_3 (\sqrt{t} + \sqrt{t_A})} \, \sin [2 k (\sqrt{t} - \sqrt{t_A})]
\end{aligned}
\end{equation}

where $0 \leq t_{\tilde{A}} \leq t_A < t$. In order to compare with the result of Barut and Duru [12] we choose as initial conditions

\begin{equation}
\phi^{(j, -)}(\textbf{k}, t_A) =  \left( \begin{array}{c} e^{- 4 i k_3 \, \sqrt{t_A}}  \\ i \, \frac{(- 1)^j \, k\, \mathrm{sign} k_3 \, - \, k_3 }{k_+}  \end{array} \right)
\end{equation}

($j = 1, 2$) and get with (23), (34) and (40) the two solutions

\begin{equation}
\phi^{(j, -)}(\textbf{k}, t) =   \left( \begin{array}{c} 
e^{- 4 i k_3 \, \sqrt{t_A}} \  e^{2 i [ (- 1)^j \, k \, \mathrm{sign} k_3 \, - \, k_3] \, (\sqrt{t} - \sqrt{t_A})}  \\ 
i \, \frac{(- 1)^j \, k\, \mathrm{ sign} k_3 \, - \, k_3 }{k_+} \ e^{2 i [ (- 1)^j \, k \, \mathrm{sign} k_3 \, + \, k_3] \, (\sqrt{t} - \sqrt{t_A})} 
\end{array} \right) .
\end{equation}

Thus, the general negative chirality bispinor solutions assume according to (13), (9) the form ($j  = 1, 2$)

\begin{equation}
\psi_{\textbf{k}}^{(j, -)}(\textbf{x},t) \ = \ c^{(j, -)}_\textbf{k} \, e^{i\textbf{k} \textbf{x}} \left( \begin{array}{c} \varphi^{(j, -)}(\textbf{k}, t)  \\ - \, \varphi^{(j, -)}(\textbf{k}, t)  \end{array} \right)
\end{equation}

with Weyl spinors

\begin{equation}
\begin{aligned}
\varphi^{(j, -)}(\textbf{k}, t) \, = \, |g(t)&|^{-1/4} \, e^{- 2 i k_3 \, (\sqrt{t_{\tilde{A}}} \, + \, \sqrt{t_A})} 
\\
& \times e^{2 i (- 1)^j \,  k \, \mathrm{sign}k_3 \, (\sqrt{t} \, - \, \sqrt{t_A})}
\left( \begin{array}{c} 
1  \\ 
i \, \frac{(- 1)^j \, k\, \mathrm{ sign} k_3 \, - \, k_3 }{k_2 + i k_1} 
\end{array} \right)
\end{aligned}
\end{equation}

and $|g| = t^3$. The positive chirality solutions can be determined by use of (14):

\begin{equation}
\psi_{\textbf{k}}^{(j, +)}(\textbf{x},t) \ = \ c^{(j, +)}_\textbf{k} \, e^{i\textbf{k} \textbf{x}} \left( \begin{array}{c} \varphi^{(j, +)}(\textbf{k}, t)  \\  \, \varphi^{(j, +)}(\textbf{k}, t)  \end{array} \right)
\end{equation}

where

\begin{equation}
\begin{aligned}
\varphi^{(j, +)}(\textbf{k}, t) = |g(t)&|^{-1/4} \, e^{ 2 i k_3 \, (\sqrt{t_{\tilde{A}}} \, + \, \sqrt{t_A})} \,
\\
& \times e^{- 2 i (- 1)^j \,  k \, \mathrm{sign}k_3 \, (\sqrt{t} \, - \, \sqrt{t_A})}
\left( \begin{array}{c} 
1  \\ 
i \, \frac{(- 1)^j \, k\, \mathrm{ sign} k_3 \, - \, k_3 }{k_2 + i k_1} 
\end{array} \right) .
\end{aligned}
\end{equation}

The four normalization constants $c^{(j, \pm)}_\textbf{k}$ are calculated with the help of (15), (16):

\begin{equation}
c^{(1, -)}_\textbf{k} \, = \, \tan \zeta \  c^{(2, -)}_\textbf{k}, \ \  c^{(2, -)}_\textbf{k} \, = \, \frac{\cos \zeta }{\sqrt{2} \, (2 \pi)^{3/2}},  \ \  c^{(j, +)}_\textbf{k} \, = \, c^{(j, -)}_\textbf{k} ,
\end{equation}

where $\cos \zeta :=   \sqrt{ (k + |k_3|)/2k}$. Appropriate linear combinations of the chirality eigenspinors yield then the solutions of Barut and Duru [12]:

\begin{equation}
\begin{aligned}
\cos \zeta \ \, \psi_{\textbf{k}} ^{(1, +)}(\textbf{x}, t) &   \ - \
 \sin \zeta \ \, \psi_{\textbf{k}}^{(2, -)}(\textbf{x}, t)  \ = 
\\
&\frac{1}{\sqrt{2} (2 \pi)^{3/2}} \, \frac{e^{i \textbf{k} \textbf{x}}}{t^{3/4}} \, 
e^{2 i \,  k \, \mathrm{sign}k_3 \, (\sqrt{t} \, - \, \sqrt{t_A})} \ 
\left( \begin{array}{c} 
0  
\\
1
\\
- \frac{\mathrm{sign} k_3}{k} \, (k_1 - i k_2)
\\
\frac{\mathrm{sign} k_3}{k} \, k_3
\end{array} \right)
\\
\\
\mathrm{sign} k_3 \,[  \sin \zeta \ \, \psi_{\textbf{k}} ^{(1, +)}(\textbf{x}, t)   & \ + \
 \cos \zeta \ \, \psi_{\textbf{k}}^{(2, -)}(\textbf{x}, t) ] \ = 
\\
&\frac{1}{\sqrt{2} (2 \pi)^{3/2}} \, \frac{e^{i \textbf{k} \textbf{x}}}{t^{3/4}} \, 
e^{2 i \,  k \, \mathrm{sign}k_3 \, (\sqrt{t} \, - \, \sqrt{t_A})} \ 
\left( \begin{array}{c} 
1
\\
0
\\
- \frac{\mathrm{sign} k_3}{k} \, k_3
\\
- \frac{\mathrm{sign} k_3}{k} \, (k_1 + i k_2)
\end{array} \right)
\\
\\
\sin \zeta \ \, \psi_{\textbf{k}} ^{2, +)}(\textbf{x}, t)   & \ - \
 \cos \zeta \ \, \psi_{\textbf{k}}^{(1, -)}(\textbf{x}, t)  \ = 
\\
&\frac{1}{\sqrt{2} (2 \pi)^{3/2}} \, \frac{e^{i \textbf{k} \textbf{x}}}{t^{3/4}} \, 
e^{- 2 i \,  k \, \mathrm{sign}k_3 \, (\sqrt{t} \, - \, \sqrt{t_A})} \ 
\left( \begin{array}{c} 
0  
\\
1
\\
\frac{\mathrm{sign} k_3}{k} \, (k_1 - i k_2)
\\
- \frac{\mathrm{sign} k_3}{k} \, k_3
\end{array} \right)
\\
\\
\mathrm{sign} k_3 \ [ \cos \zeta \ \, \psi_{\textbf{k}} ^{(2, +)}(\textbf{x}, t)   & \ + \ \sin \zeta \ \, \psi_{\textbf{k}}^{(1, -)}(\textbf{x}, t) ] \ = 
\\
&\frac{1}{\sqrt{2} (2 \pi)^{3/2}} \, \frac{e^{i \textbf{k} \textbf{x}}}{t^{3/4}} \, 
e^{- 2 i \,  k \, \mathrm{sign}k_3 \, (\sqrt{t} \, - \, \sqrt{t_A})} \ 
\left( \begin{array}{c} 
1
\\
0
\\
\frac{\mathrm{sign} k_3}{k} \, k_3
\\
\frac{\mathrm{sign} k_3}{k} \, (k_1 + i k_2)
\end{array} \right).
\end{aligned}
\end{equation}
\\
In the second case we consider exact solutions of Einstein`s field equations when the material content is described by the "perfect fluid" stress tensor $T_{\alpha \beta} = ({\cal{E}} + {\cal{P})} u_{\alpha} u_{\beta} \, + \, {\cal{P}} g_{\alpha \beta}$, where ${\cal{P}}$ denotes the pressure, ${\cal{E}}$ the energy density plus total internal energy, and $u_{\alpha}$ the four velocity. The equation of states reads ${\cal{P}} = (\gamma - 1) {\cal{E}}$ with $\gamma$ a constant. $\gamma = 2$ describes the so-called "stiff matter" model. A special case is given by the line element (17) with $\mu = 1,\ \ \nu =  1/2$ [13].

We begin with (10), (11) and (19) and investigate again only the case with negative chirality:

\begin{equation}
\begin{aligned}
\partial_t \phi_1^{(j, -)}(\textbf{k}, t) &= \frac{k_+}{\sqrt{t}} \, e^{- 2i k_3 t} \, \phi_2^{(j, -)}(\textbf{k}, t)
\\
\partial_t \phi_2^{(j, -)}(\textbf{k}, t) &= - \frac{k_-}{\sqrt{t}} \, e^{ 2i k_3 t} \, \phi_1^{(j, -)}(\textbf{k}, t). 
\end{aligned}
\end{equation}

Exact solutions of this system are given by\footnote{A similar result has been found in ref. [14] for the first components of the positive chirality solutions, $\phi_1^{(j, +)}$, which are related to (50), (51) via (14).}

\begin{equation}
\begin{aligned}
\phi_1^{(1, -)}(\textbf{k}, t) \, = \,& (- 2i k_3 t)^{- 1/4} \, e^{- ik_3 t} \, W_{- \frac{1}{4} - i \eta; \frac{1}{4}}(2i k_3 t)
\\
\phi_2^{(1, -)}(\textbf{k}, t) \, = \,& \frac{\sqrt{- 2i k_3}}{k_+} \, (- 2i k_3 t)^{- 3/4} \, e^{ ik_3 t} \, 
\\
& \times [i \eta \, W_{- \frac{1}{4} - i \eta; \frac{1}{4}}(2i k_3 t) \ - \  W_{ \frac{3}{4} - i \eta; \frac{1}{4}}(2i k_3 t) ]
\end{aligned}
\end{equation}

and

\begin{equation}
\begin{aligned}
\phi_1^{(2, -)}(\textbf{k}, t) \, = \,& (- 2i k_3 t)^{- 1/4} \, e^{- ik_3 t} \, W_{ \frac{1}{4} + i \eta; \frac{1}{4}}(- 2i k_3 t)
\\
\phi_2^{(2, -)}(\textbf{k}, t) \, = \,& i \, \frac{k_- \, \mathrm{sign}k_3}{\sqrt{ 2i k_3}} \, (- 2i k_3 t)^{- 3/4} \, e^{ ik_3 t} \, 
\\
& \times [ W_{ \frac{1}{4} + i \eta; \frac{1}{4}}(- 2i k_3 t) \ + \ (i \eta - \frac{1}{2}) \, W_{ - \frac{3}{4} +  i \eta; \frac{1}{4}}(- 2i k_3 t) ]
\end{aligned}
\end{equation}

[15] where $W_{\mu; \nu}(z)$ denotes Whittaker`s function [11], and we defined with (27)

\begin{equation}
\eta := \mathrm{sign} k_3 \, \eta_{\frac{1}{2}} \equiv  \kappa^2/2k_3.
\end{equation}

 The bispinor solutions read owing to (13) with (9)

\begin{equation}
\begin{aligned}
\psi^{(1, -)}(\textbf{x},t) \, = \, & c^{(1, -)}_\textbf{k} \, e^{i\textbf{k} \textbf{x}} \, |g(t)|^{- 1/4}
\left( \begin{array}{c}  e^{ik_3t}  \, \phi_1^{(1, -)}(\textbf{k}, t)
\\
 e^{- ik_3t}  \, \phi_2^{(1, -)}(\textbf{k}, t)
\\
- e^{ik_3t}  \, \phi_1^{(1, -)}(\textbf{k}, t)
\\
- e^{- ik_3t}  \, \phi_2^{(1, -)}(\textbf{k}, t)
\end{array} \right)
\\
\\
\psi^{(2, -)}(\textbf{x},t) \, = \, & c^{(2, -)}_\textbf{k} \, e^{i\textbf{k} \textbf{x}} \, |g(t)|^{- 1/4}
\left( \begin{array}{c}   e^{ik_3t}  \, \phi_1^{(2, -)}(\textbf{k}, t)
\\
e^{- ik_3t}  \, \phi_2^{(2, -)}(\textbf{k}, t)
\\
- e^{ik_3t}  \, \phi_1^{(2, -)}(\textbf{k}, t)
\\
- e^{- ik_3t}  \, \phi_2^{(2, -)}(\textbf{k}, t)
\end{array} \right).
\end{aligned}
\end{equation}

Here, $|g(t)| = t^2$. The positive chirality solutions are then by means of (14) found to be

\begin{equation}
\begin{aligned}
&\psi^{(1, +)}(\textbf{x},t) = c^{(1, +)}_\textbf{k} \, \frac{e^{i\textbf{k} \textbf{x}}}{\sqrt{t}}
\\
& \times \left( \begin{array}{c}  ( 2i k_3 t)^{- 1/4} \,  W_{- \frac{1}{4} + i \eta; \frac{1}{4}}(- 2i k_3 t) 
\\
\frac{\sqrt{ 2i k_3}}{k_+} \, ( 2i k_3 t)^{- 3/4} \,  [i \eta \, W_{- \frac{1}{4} + i \eta; \frac{1}{4}}(- 2i k_3 t) \ + \  W_{ \frac{3}{4} + i \eta; \frac{1}{4}}(- 2i k_3 t) ] 
\\
 ( 2i k_3 t)^{- 1/4} \,  W_{- \frac{1}{4} + i \eta; \frac{1}{4}}(- 2i k_3 t)
\\
\frac{\sqrt{ 2i k_3}}{k_+} \, ( 2i k_3 t)^{- 3/4} \,  [i \eta \, W_{- \frac{1}{4} + i \eta; \frac{1}{4}}(- 2i k_3 t) \ + \  W_{ \frac{3}{4} + i \eta; \frac{1}{4}}(- 2i k_3 t) ]
\end{array} \right)
\\
\\
&\psi^{(2, +)}(\textbf{x},t) = c^{(2, +)}_\textbf{k} \, \frac{e^{i\textbf{k} \textbf{x}}}{\sqrt{t}}
\\
& \times \left( \begin{array}{c}  ( 2i k_3 t)^{- 1/4} \,  W_{ \frac{1}{4} - i \eta; \frac{1}{4}}( 2i k_3 t) 
\\  i \, \frac{k_- \, \mathrm{sign}k_3}{\sqrt{- 2i k_3}} \, ( 2i k_3 t)^{- 3/4} \, [ W_{ \frac{1}{4} - i \eta; \frac{1}{4}}( 2i k_3 t) \ - \ (i \eta + \frac{1}{2}) \, W_{ - \frac{3}{4} -  i \eta; \frac{1}{4}}( 2i k_3 t) ]
\\
( 2i k_3 t)^{- 1/4} \,  W_{ \frac{1}{4} - i \eta; \frac{1}{4}}( 2i k_3 t)
\\
i \, \frac{k_- \, \mathrm{sign}k_3}{\sqrt{- 2i k_3}} \, ( 2i k_3 t)^{- 3/4} \, [ W_{ \frac{1}{4} - i \eta; \frac{1}{4}}( 2i k_3 t) \ - \ (i \eta + \frac{1}{2}) \, W_{ - \frac{3}{4} -  i \eta; \frac{1}{4}}( 2i k_3 t) ]
\end{array} \right).
\end{aligned}
\end{equation}

In order to determine the short - and long - time behavior of these exact solutions it suffices to consider $\phi^{(j, -)}(\textbf{x},t)$ as given by (50) and (51). For $t \ll |k_3|^{- 1}$ (where for the sake of simplicity we put for the lower limit of the integral in eq. (12): $t_{\tilde{A}} \equiv 0$) one gets

\begin{equation}
\begin{aligned}
\phi^{(1, -)}(\textbf{k},t) \, = \, & \sqrt{\pi} \, e^{i \frac{\pi}{4} \mathrm{sign} k_3} \, 
\left( \begin{array}{c}  \frac{1}{\Gamma(1 + i \eta)} \, - \, \frac{2 \sqrt{2i k_3 t}}{\Gamma(\frac{1}{2} + i \eta)}
\\
- \frac{\sqrt{2i k_3}}{k_2 + i k_1} \left[ \frac{1}{\Gamma(\frac{1}{2} + i \eta)} \ - \ \frac{2 \sqrt{2i k_3 t}}{\Gamma(i \eta)} \right]
\end{array} \right) \ + \ O(k_3t)
\\
\phi^{(2, -)}(\textbf{k},t) \, = \, & \sqrt{\pi} \, 
\left( \begin{array}{c}  \frac{1}{\Gamma(\frac{1}{2} - i \eta)} \, - \, \frac{2 \sqrt{- 2i k_3 t}}{\Gamma (- i \eta) }
\\
i \, \frac{(k_2 - i k_1) \, \mathrm{sign} k_3 }{\sqrt{2i k_3}} \frac{1}{\Gamma(1 - i \eta)} \left[1 \ - \ \frac{ 2 \sqrt{- 2i k_3 t} \  \Gamma(1 - i \eta)}{\Gamma(\frac{1}{2} - i \eta)}  \right]
\end{array} \right)
\\
& \ + \ O(k_3t)
\end{aligned}
\end{equation}

whereas in the asymptotic regime $t \gg |k_3|^{- 1}$ holds

\begin{equation}
\begin{aligned}
\phi^{(1, -)}(\textbf{k},t) \, = \, &  \, e^{i \frac{\pi}{4} \mathrm{sign} k_3} \, (2i k_3t)^{- i\eta} 
\left( \begin{array}{c}  \ \frac{e^{- 2i k_3t}}{\sqrt{2i k_3t}} 
\\
- \frac{\sqrt{2i k_3}}{k_2 + ik_1} 
\end{array} \right) \left( 1 +  O[(k_3 t)^{- 1}]  \right)
\\
\phi^{(2, -)}(\textbf{k},t) \, = \, & \, (- 2i k_3t)^{ i\eta}
\left( \begin{array}{c}  1 
\\
i \, \frac{(k_2 - ik_1) \, \mathrm{sign} k_3}{\sqrt{2i k_3}} \frac{e^{2i k_3t}}{\sqrt{- 2i k_3t}}
\end{array} \right)  \left( 1 +  O[(k_3 t)^{- 1}]  \right)
\end{aligned}
\end{equation}

Let us compare this outcome with the corresponding behavior of our solution (34). The time - evolution operator reads

\begin{equation}
K_{\textbf{k}}^{(-)}(t| t_A) \, = \,
\left( \begin{array}{rr} \left( K_{\textbf{k}}^{(-)}(t|t_A) \right)_{11}     &  \left( K_{\textbf{k}}^{(-)}(t|t_A) \right)_{12}  \\ 
\\  - \left( K_{\textbf{k}}^{(-)}(t|t_A) \right)^{\ast}_{12}  & \left( K_{\textbf{k}}^{(-)}(t|t_A) \right)_{11}^{\ast}    \end{array}  \right) .
\end{equation}

In the asymptotic limit $|k_3|t \rightarrow \infty$  one gets after some algebra for $t_A = 0$

\begin{equation}
\begin{aligned}
\left( K_{\textbf{k}}^{(-)}(t|0) \right)_{11} \,  = & \, 1 + O( \eta) + \, \sqrt{\pi} \, |\eta |  \, \frac{e^{- 2i k_3t}}{\sqrt{ 2i k_3t}}  ,
\\
\left( K_{\textbf{k}}^{(-)}(t|0) \right)_{12} \,  = & \, \frac{k_2 + ik_1}{\sqrt{- 2i k_3}} \, \{
- i \sqrt{\pi} \, \mathrm{sign} k_3 \, + \, \frac{e^{- 2i k_3t}}{\sqrt{- 2i k_3t}} \, [1 + \, O(\eta)] \} .
\end{aligned}
\end{equation}

According to (23) the asymptotic  expansion of $\phi^{(j, -)}(\textbf{k},t)$ can be formally written as

\begin{equation}
\phi^{(j, -)}(\textbf{k},t) \, \sim \, K_{\textbf{k}}^{(-)}(t|0) \, \phi^{(j, -)}(\textbf{k},0)
\end{equation}

where we choose as initial condition the exact results (55) at $t = 0$:

\begin{equation}
\begin{aligned}
\phi^{(1, -)}(\textbf{k},0) \, = \, & \sqrt{\pi} \, e^{i \frac{\pi}{4} \mathrm{sign} k_3} \, 
\left( \begin{array}{c}  \frac{1}{\Gamma(1 + i \eta)}
\\
- \frac{\sqrt{2i k_3}}{k_2 + i k_1} \, \frac{1}{\Gamma(\frac{1}{2} + i \eta)} 
\end{array} \right)
\\
\phi^{(2, -)}(\textbf{k},0) \, = \, & \sqrt{\pi} \, 
\left( \begin{array}{c}  \frac{1}{\Gamma(\frac{1}{2} - i \eta)}
\\
i \, \frac{(k_2 - i k_1) \, \mathrm{sign} k_3 }{\sqrt{2i k_3}} \frac{1}{\Gamma(1 - i \eta)} 
\end{array} \right) .
\end{aligned}
\end{equation}

Eq. (25) implies that the asymptotic regime is established when the condition $2 |k_3|  s(t) / \mu \gg 1$ is satisfied. This means that for fixed $s$ the asymptotic behavior can also be obtained by letting $|k_3|$ tend to infinity. As a consequence constant terms of order $ \eta \sim k_3^{- 1}$ \ can be neglected against unity in the expressions (58) and (60) so that one gets with (59)

\begin{equation}
\begin{aligned}
\phi^{(1, -)}(\textbf{k},t) \, & \sim \, e^{i \frac{\pi}{4} \mathrm{sign} k_3}  \,
\left( \begin{array}{c}
\frac{e^{- 2i k_3t}}{\sqrt{2i k_3t}} 
\\
- \frac{\sqrt{2i k_3}}{k_2 + ik_1} \, 
\end{array} \right)
\\
\phi^{(2, -)} (\textbf{k},t) \, & \sim \,  
\left( \begin{array}{c} 
1 
\\
i \, \frac{(k_2 - ik_1) \, \mathrm{sign} k_3} {\sqrt{2i k_3}} \,  \frac{e^{2i k_3t}}{\sqrt{- 2i k_3t}}  
\end{array} \right) .
\end{aligned}
\end{equation}

The comparison with the asymptotic expansions (56) of the exact results shows agreement apart from the factors $(\pm 2i k_3t)^{\mp i \eta}$. However, these are up to a constant given by  a time dependent phase  factor with vanishing frequency for large times, $\omega(t) = \kappa^2 ln(2 |k_3|t)/2 |k_3| t$, and can therefore be disregarded in the asymptotic limit.

In the short time limit $t \rightarrow 0 \ (t_{\tilde{A}} = t_A = 0)$ one obtains from eq.s (37), (38)

\begin{equation}
\begin{aligned}
R(z; s) \, &= \, \frac{4}{\pi} \, \sinh[(1 - \delta) \, z/2] \ [1 + O(s^{2 \delta})] \nonumber
\\
Z(z; s) \, &= \, \frac{2 \, e^{- (1 - \delta) z/2}}{\pi x(s)} \ [1 + O(s^{2 \delta})]
\end{aligned}
\end{equation}

so that with (57) and (34) the negative chirality time evolution operator assumes the simple form

\begin{equation}
K_{\textbf{k}}^{(-)}(t| 0) \, = \,
\left( \begin{array}{rr} 1 \ \ \   &  \frac{k_+}{\mu \, \delta} \, s^{\delta}  \\ -  \frac{k_-}{\mu \, \delta} \, s^{\delta}   &  1 \ \ \   \end{array} \right) \, [1 + O(s^{2 \delta})] .
\end{equation}

Insertion of (60), (62) into (23) then yields for $\delta = 1/2$ the correct short - time limits (55).
\\
\section*{IV. CONCLUSION}
Specializing to axisymmetric anisotropic BI backgrounds described by the  line - element (17), we derived with eq.s (34) - (38) a general expression of the time evolution operator  for fields obeying the Dirac - Weyl equations. It has been shown that the use of approximation (29) is reasonable. This approximation leads to an analytical result rendering the exact short - and large - time behavior of massless fermions which propagate in spacetimes (17) satisfying $1 - \mu \leq \nu$ (i. e. with (22): $\delta \leq 1$). The comparison with an exactly solvable anisotropic stiff fluid model confirms the validity of eq.s (34) - (38).

An advantage of the computation of the time evolution operator consists in that the initial value problem has been automatically taken care of. Conversely, given any asymptotic solution one can reconstruct the pertaining initial conditions at time $t_A$. Hence, if one can not dispose of exact solutions of the Dirac - Weyl system, which is the case for anisotropic backgrounds (17) (with only a few exceptions to the rule, e.g. the above mentioned stiff - fluid model), this approach provides approximate analytical solutions which come as close as possible to the exact solutions.

By means of a  $1 - \delta \, -$ expansion of eq.s (34) - (38) it is possible to study small perturbations about flat RW backgrounds (the case $\delta = 1$), and it has been explicitly shown that the limit $\delta \rightarrow 1$ yields the exact result for that flat RW background which satisfies $\nu = 1/2 = \mu$. However, the case $\delta = 1$ is special because all spacetimes fulfilling this condition are conformally equivalent (which is no longer true for $\delta < 1$). As a consequence, the model with $\nu= 1/2 = \mu$ already represents the general result.
\\
\section*{ACKNOWLEDGMENTS}
I thank K. - H. - Lotze and A. Wipf for interesting discussions and helpful comments.
\newpage

[1]\ \ \  D. R. Brill and J. A. Wheeler, Rev. Mod. Phys. \textbf{29}, 465 (1957).

[2]\ \ \  L. Parker, Phys. Rev. D \textbf{3}, 346 (1971).

[3]\ \ \  Ya. B. Zel`dovich and A. A. Starobinski, Zh. Eksp. Zeor. Fiz. \textbf{26},

\ \ \ \ \ \ \,373 (1977); see also N. D. Birrell and P. C. W. Davies, J. Phys. A

\ \ \ \ \ \ \,\textbf{13}, 2109 (1980), and K. H. Lotze, Class. Quantum Grav. \textbf{3}, 81

\ \ \ \ \ \ \,(1986).

[4]\ \ \  Ya. B. Zel`dovich, Sov. Phys. - JETP \textbf{12}, 307 (1970).

[5]\ \ \  See e.g. C. Pitrou, T.S. Pereira, and J. P. Uzan, J. Cosmol. Astro-

\ \ \ \ \ \ \,part. Phys. \textbf{04} (2008) 004; H. C. Kim and M. Minamitsuji, Phys.

\ \ \ \ \ \ \,Rev. D \textbf{81}, 083517 (2010).

[6]\ \ \  M. Henneaux, Phys. Rev. D \textbf{21}, 857 (1980).

[7]\ \ \  B. Saha and T. Boyadjiev, Phys. Rev. D \textbf{69}, 124010 (2004).

\ \ \ \ \ \ \,Phys. Lett. A \textbf{128}, 25 (1988).

[8]\ \ \  M. Wollensak, J. Math. Phys. \textbf{39}, 5934 (1998).

[9]\ \ \ C. W. Misner, K. S. Thorne, and J. A. Wheeler, $Gravitation$

\ \ \ \ \ \ \,(Freeman, New York, 1973).

[10]\ \,R. U. Sexl and H. K. Urbantke, $Gravitation \, und \, Kosmologie$, 3rd

\ \ \ \ \ \ \,ed. (BI-Wiss.-Verl., Mannheim, 1987).

[11]\ \,M. Abramowitz and I. A. Stegun, $Pocketbook \, of Mathematical \, Functions$,

\ \ \ \ \ \ \,(H. Deutsch, Frankfurt, 1984).

[12]\ \,A. O. Barut and I. H. Duru, Phys. Rev. D \textbf{36}, 3705 (1987).

[13]\ \,D. Kramer, H. Stephani, M. MacCallum and E. Herlt, $Exact \, Solutions$

\ \ \ \ \ \ \,$of \,Einstein`s \, Field \, Equations$ (Deutscher Verlag d. Wiss., Berlin,

\ \ \ \ \ \ \,1980).

[14]\ \,L. O. Pimentel, Int. J. Theor. Phys. \textbf{32}, 979 (1993).

[15]\ \,E. Kamke, $Differentialgleichungen: \, L \ddot{o} sungsmethoden \, und \, L \ddot{o} sungen$

\ \ \ \ \ \ \,9th ed. (Teubner, Stuttgart, 1977)

\end{document}